\documentclass[aps,pre,
 amsmath,amssymb,twocolumn,showpacs,showkeys,superscriptaddress
]{revtex4-1}

\usepackage{graphicx}
\usepackage{dcolumn}
\usepackage{bm}
\usepackage{color}

\begin{document}

\preprint{AIP/123-QED}

\title[Sessile and pendant drops excited by SAWs]{Dynamics of sessile and pendant drop excited by surface acoustic waves: gravity effects and correlation between oscillatory and translational motions.}

\author{A. Bussonni\`{e}re}
\affiliation{Universit\'{e} Lille 1, International Laboratory LEMAC/LICS, IEMN, UMR CNRS 8520, Avenue Poincar\'e, 59652 Villeneuve d'Ascq, France
}%
\author{M. Baudoin}
 \email{michael.baudoin@univ-lille1.fr}
  \homepage{http://films-lab.univ-lille1.fr}
\affiliation{Universit\'{e} Lille 1, International Laboratory LEMAC/LICS, IEMN, UMR CNRS 8520, Avenue Poincar\'e, 59652 Villeneuve d'Ascq, France
}%
\author{P. Brunet}
\affiliation{Laboratoire Mati\`{e}re et Syst\`{e}mes Complexes, UMR CNRS 7057, Universit\'{e} Paris Diderot, 10 rue Alice Domon et L\'{e}onie Duquet, 75205 Paris cedex 13, France
}%

\author{O. Bou Matar}
\affiliation{Universit\'{e} Lille 1, International Laboratory LEMAC/LICS, IEMN, UMR CNRS 8520, Avenue Poincar\'e, 59652 Villeneuve d'Ascq, France
}%

\date{\today}

\begin{abstract}
When sessile droplets are excited by ultrasonic traveling surface acoustic waves (SAWs), they undergo complex dynamics with both oscillations and translational motion. While the nature of the Rayleigh-Lamb quadrupolar drop oscillations has been identified, their origin and their influence on the drop mobility remains unexplained. Indeed the physics behind this peculiar dynamics is complex with nonlinearities involved both at the excitation level (acoustic streaming and radiation pressure) and in the droplet response (nonlinear oscillations and contact line dynamics). In this paper, we investigate the dynamics of sessile and pendant drops excited by SAWs. For pendant drops, so-far unreported dynamics are observed close to the drop detachment threshold with the suppression of the translational motion. Away from this threshold, the comparison between pendant and sessile drop dynamics allows us to identify the role played by gravity or more generally by an initial or dynamically induced stretching of the drop. In turn, we elucidate the origin of the resonance frequency shift, as well as the origin of the strong correlation between oscillatory and translational motion. We show that for sessile drops, the  velocity is mainly determined by the amplitude of oscillation and that the saturation observed is due to the nonlinear dependence of the drop response frequency on the dynamically induced stretching. \end{abstract}

\maketitle

\section{\label{sec:i}Introduction}

Surface acoustic waves (SAWs) are versatile tools for the actuation of fluids at small scales. In digital microfluidic, they can be used to move \cite{abc_wixforth_2004, saa_renaudin_2006, pre_brunet_2010}, divide \cite{loc_collignon_2015}, merge \cite{mn_babetta_2012}, atomize \cite{ieee_shiokawa_1989,jjap_shiokawa_1990, jjap_chono_2004,pof_qi_2008, prl_tan_2009,pre_collins_2012}, mix \cite{prl_frommelt_2008} or heat \cite{ieeeus_kondoh_2005,jjap_ito_2007,saa_kondoh_2009,pnas_reboud_2012, ieeetuffc_rouxmarchand_2015,afm_shilton_2015} sessile droplets. In microchannels, they can induce fluid pumping \cite{loc_girardo_2008,apl_cecchini_2008} and mixing \cite{jmm_tseng_2006, epl_friend_2009,loc_dentry_2014}. Finally, in both configurations, they can be used to manipulate and sort particles \cite{jap_shilton_2008,loc_huang_2008,lc_huang_2009,mn_raghavan_2010,apl_tran_2012,pnas_ding_2012,ieee_guo_2014,apl_collins_2014} and cells \cite{loc_shi_2009,loc_franke_2010,loc_hartmann_2014,loc_bussoniere_2014,apl_sivanatha_2014,pnas_ding_2014,pnas_li_2015,pnas_guo_2015}. More recently, new types of surface acoustic waves, the so called swirling waves (2D counterparts of Bessel beams) \cite{prap_riaud_2015,pre_riaud_2015}, have been under investigation to achieve new operations, such as 3D on-chip single particle manipulation \cite{prl_baresch_2015} or vortical flow synthesis with controlled topology \cite{pre_riaud_2014}. 

\begin{figure}[htbp]
\centerline{\includegraphics[width=0.45 \textwidth]{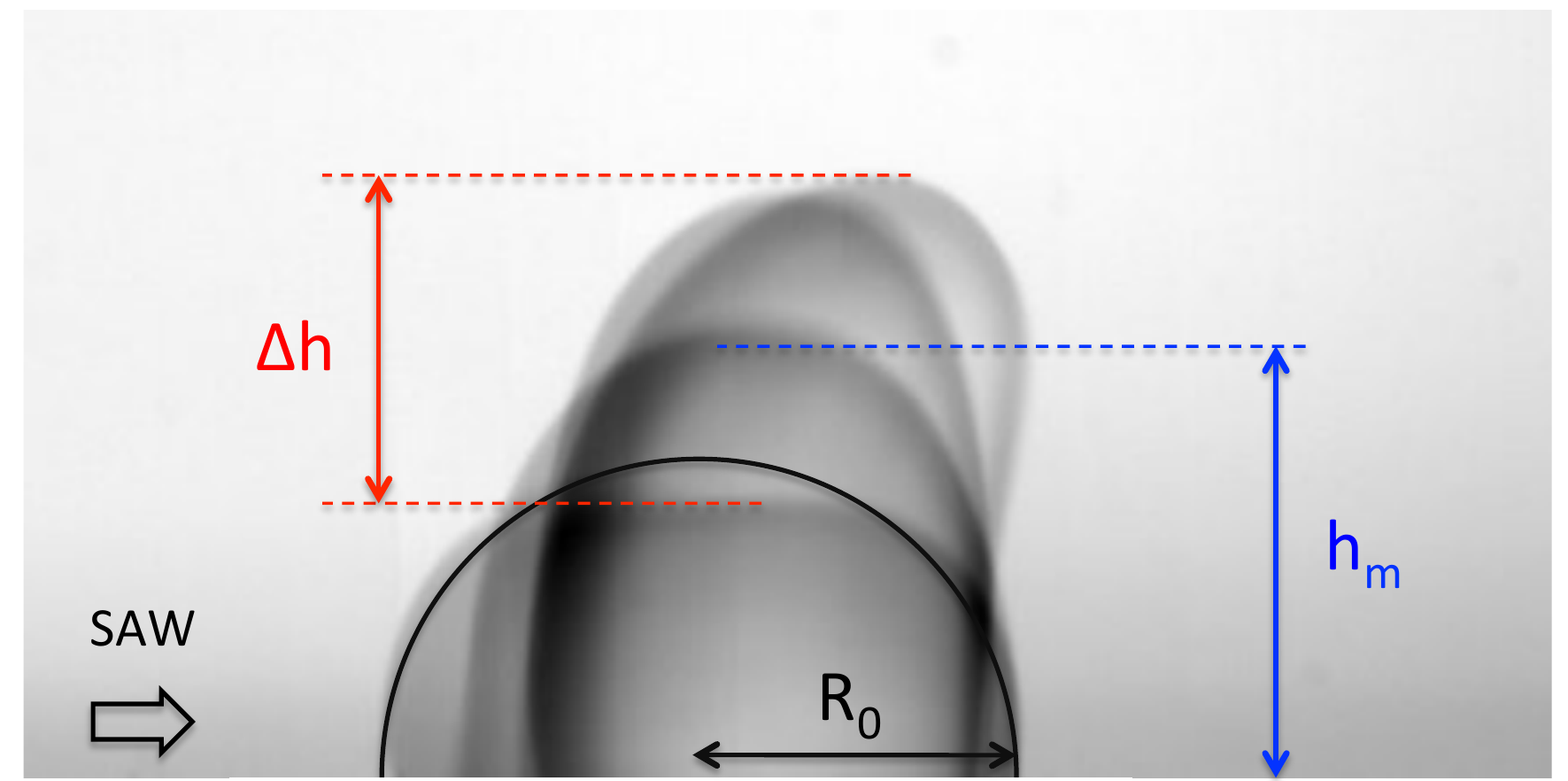}}
\caption{\label{fig:def} Successive deformations of a drop excited by a Rayleigh SAW, leading to both translational motion at velocity $V$ and oscillations of amplitude $\Delta h$ and frequency $f_\text{osc}$. The radiated acoustic wave leads to an average drop height $h_m$. The plain line circle and base radius $R_0$ correspond to the shape without acoustic forcing.}
\end{figure}

However, despite an extensive literature on the use of SAW for labs-on-chips actuation \cite{b_alavarez_2009,rmp_friend_2011,loc_ding_2013}, a clear understanding of the underlying physics is still missing for many of these systems \cite{arfm_yeo_2014}. One of the reason is that the nonlinear coupling between the acoustic waves and the liquid response involves time and length scales that differ by several order of magnitude, along with nonlinear effects, which render the analysis and simulation of these behaviors extremely difficult. A difficult problem is the physical understanding of droplets dynamics excited by planar propagative SAW. In this case the droplet undergoes both oscillatory and translational motion \cite{sab_beyssen_2006,pre_brunet_2010} that are strongly coupled to each other \cite{apl_baudoin_2012}. The droplet vibrations have been identified \cite{pre_brunet_2010} to be inertio-capillary Rayleigh-Lamb quadrupolar oscillations, whose frequency $f_{\text{osc}}$ is roughly 100 Hz, while the acoustic excitation is around $f_{\text{saw}}$ = 20 MHz. However the origin of these oscillations \cite{l_blamey_2013,arfm_yeo_2014} and their influence on the droplet translational motion \cite{apl_baudoin_2012} remain unexplained. As the SAW is radiated within the drop, it can generate both Eckart acoustic streaming and radiation pressure at the liquid-air interface, and the relative contribution of both effects depends on the attenuation length of the wave in the liquid \cite{pre_brunet_2010}. Numerical simulations showed that for water and $f_\text{saw}$ = 20 MHz, bulk attenuation is weak enough to allow for significant acoustic pressure to reach the interface \cite{pre_brunet_2010}. The radiation pressure acts by pushing the drop free-surface upwards, which leads to an average drop deformation and oscillations even at moderate acoustic power. Figure \ref{fig:def} shows a typical sequence for the dynamics for the drop, together with the geometrical definitions.

In this paper, we investigate the dynamics of sessile and pendant drops excited by Rayleigh-type SAW of frequency around 20 MHz. It is shown that even for relatively small droplets, gravity strongly affects the drop dynamics. For pendant drops, new regimes are observed close to the detachment threshold with the appearance of a quasi static equilibrium. Away from this threshold, the comparison of sessile and pendant drops dynamics allows us to identify the role played by gravity on the frequency $f_{\text{osc}}$ and amplitude of oscillations $\Delta h$: since drops are nonlinear oscillators, their characteristic frequency depends on the average stretching of the string (here the drop shape). In the case of pendant drop, this stretching is mainly induced by stationary effects (gravity and radiation pressure) that both act in the same direction. For sessile drops however, gravity and radiation pressure act in opposition and the average stretching is mainly induced by nonlinear dynamical effects, which depend on $\Delta h$. Thus in this case $f_{\text{osc}}$ is strongly influenced by $\Delta h$.

The motion of the contact line over one cycle is then analyzed for both pendant and sessile droplets. While the contact line dynamics is shown to  depend also on gravity, it essentially remains a linear function of the amplitude of oscillations in both cases. As a consequence, both the frequency $f_{\text{osc}}$ and the contact line motion over one cycle are directly dependent on the drop amplitude of oscillations for sessile droplets. In this case, it is thus possible to determine a relation between the droplet velocity $V$ and the amplitude $\Delta h$. This equation quantitatively compares to the experimentally observed tendencies for all drop sizes, and in particular the saturation of the droplet velocity as a function of the amplitude of oscillation, previously observed in ref [\onlinecite{apl_baudoin_2012}]. Consequently, it is shown that this saturation is mainly induced by the decrease of $f_{\text{osc}}$ with $\Delta h$.

In section II, the experimental setup is presented and the relevant dimensionless numbers are introduced. In section III, we first analyze the motion and oscillation of both sessile and pendant drops and compare them in order to determine the effect of gravity. Finally, in section IV the correlation between the droplet oscillations and translational motion is interpreted in light of previous results.

\section{\label{sec:es}Method}

\subsection{\label{ss:rdn}Description of the setup}

\begin{figure}[htbp]
\centerline{\includegraphics[width=0.5 \textwidth]{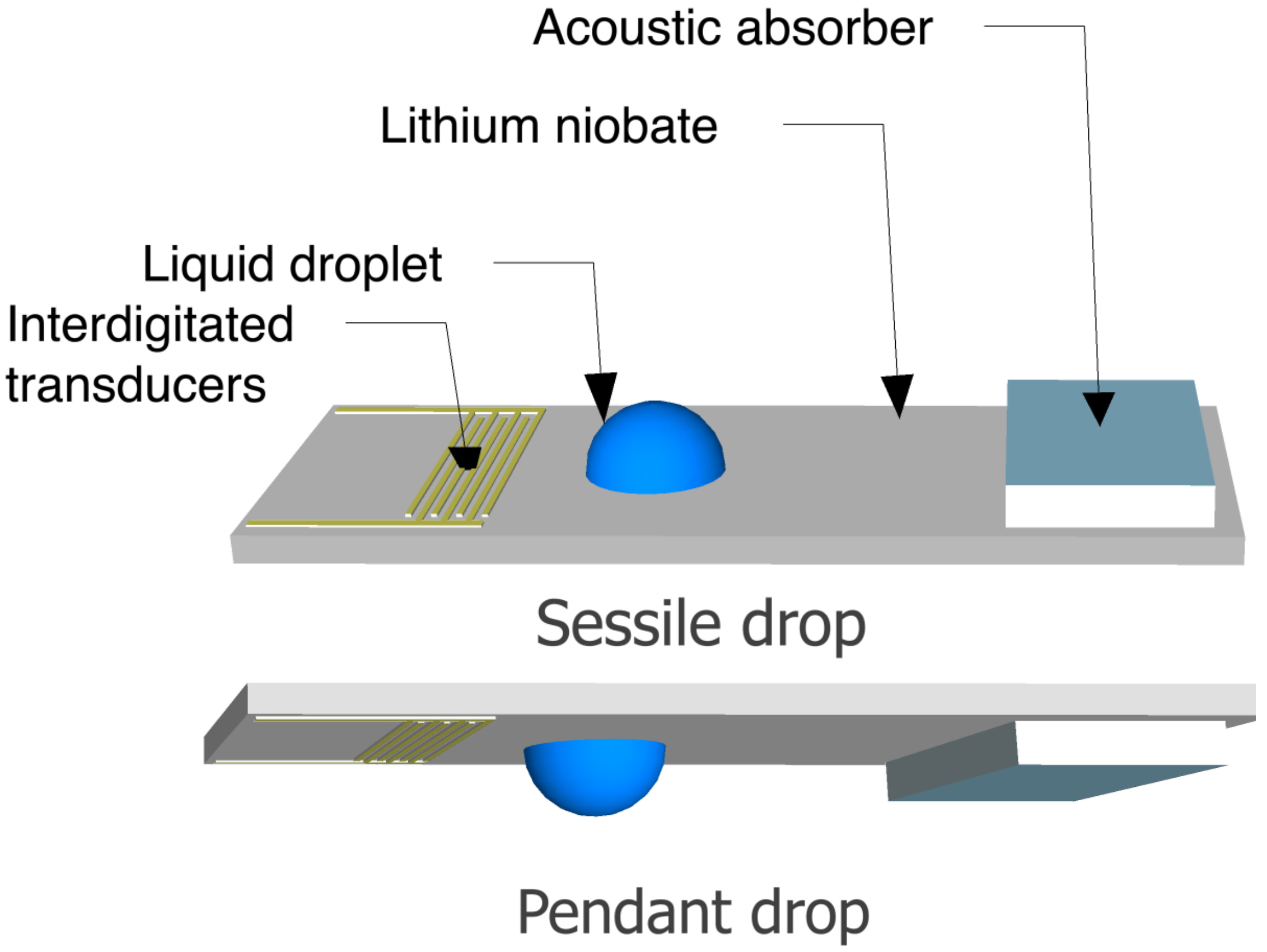}}
\caption{\label{fig:setup} Sketch of the experimental setup}
\end{figure}

Sessile and pendant drop are actuated by Rayleigh-type surface acoustic waves at driving frequency $20$ MHz synthesized at the surface of a  1.05 mm thick (Y-Cut 128$^{\circ}$) Lithium Niobate (LiNbO3) piezoelectric surface by interdigitated transducers (IDTs). The latter are excited by a high frequency generator (IFR 2023A) and a Empower RF 1037 amplifier (see Fig. \ref{fig:setup}). The IDTs have been fabricated by successively sputtering a titanium (Ti) layer (20 nm thick) and a gold (Au) layer (200 nm thick) on the LiNbO$_3$ substrate. Spacing and width of the interdigitated fingers (both equal to $a$ = 43.75 $\mu$m in the present system) determine the frequency of the surface acoustic wave according to the law $f_{\text{saw}} = c_s / \lambda = c_s / 4a$ where $\lambda$ is the wavelength and $c_s \approx$ 3484 m.s$^{-1}$ is the sound speed in the substrate in the z-direction. The amplitude of the acoustic wave was measured with a Mach-Zender inferterometer. The substrate surface was treated by a self-assembled monolayer (SAM) of OTS (octadecyltricholorisilane) making it hydrophobic (static contact angle of 98$^\circ$) and with weak contact angle hysteresis (15$^\circ$).

In sessile drop experiments, a droplet of calibrated volume is deposited on the top surface of the substrate corresponding to the active one. In pendant drop experiments, the setup is put upside down. Then waves are emitted and the droplet dynamics is recorded with a high speed camera (Photron SA3) with appropriate optics to obtain a close enough magnification of the drop. Finally, the drop oscillations and displacement, as well as the contact line dynamics are analyzed with ImageJ software (see Movies 1 and 2 in SI illustrating the typical dynamic of a sessile and a pendant drop).

\subsection{\label{ss:rdn}Relevant dimensionless numbers}

The relevant dimensionless number in this study are the Bond number $Bo$ and the acoustical Weber number $We_\text{ac}$.

\begin{figure}[htbp]
\centerline{\includegraphics[width=0.45 \textwidth]{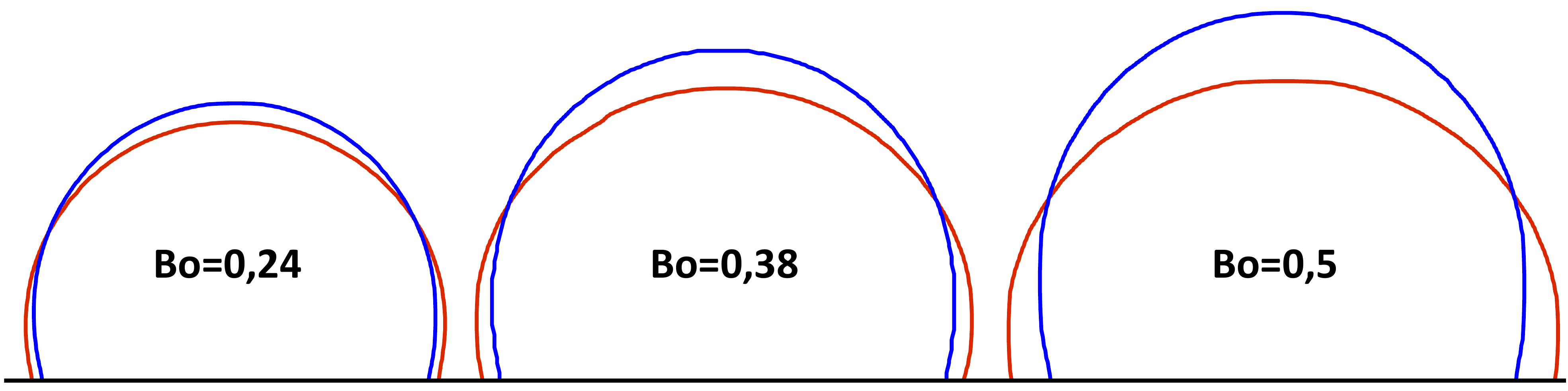}}
\caption{\label{fig:bond} Comparison of the initial static shape of sessile (red) and pendant (blue) drops of volume $5 \; \mu l$, $10 \; \mu l$ and $15 \; \mu l$ corresponding to Bond numbers of $0.24$, $0.38$ and $0.5$ respectively. }
\end{figure}

The Bond number $Bo = \rho_l g L^2 / \sigma$ compares capillary effects to gravity effects, with $\rho_l$ the density, $g$ the standard gravity, L the characteristic size of the system and $\sigma$ the surface tension. In the case of a droplet, the characteristic length $L$ corresponds to the radius of the drop at rest $R_o$ and the Bond number quantifies the ability of gravity to deform droplets (departure from spherical shape imposed by surface tension). In the present experiments, we considered droplets of $2$, $5$, $10$ and $15 \mu l$ corresponding to radii of $R_o \approx (3 V / 2 \pi )^{1/3}$ (if we consider that the drop are hemispherical) and thus Bo of respectively $0.13$, $0.24$, $0.38$ and $0.5$. Fig. \ref{fig:bond} compares the shape of these drops at rest when they are pendant and sessile corresponding to the reversal of the gravity field.  For the four volumes considered, the relative deformation spans between $4.5 \%$ and $18 \%$ of the drop initial height, giving a direct measurement of the static effect of gravity for the four drop volumes considered.


The acoustical Weber number $We_\text{ac}$ compares the acoustic radiation pressure  to the capillary effects at the drop surface. This number, initially introduced in the study of acoustic levitation  \cite{pof_shi_1995} or acoustically induced atomization \cite{pre_collins_2012}, characterizes the ability of acoustic waves to deform interfaces, which shape is maintained by surface tension. For a droplet excited by SAWs, the air-water interface is almost a perfect mirror for the wave and the radiation pressure $p_r$ can be estimated (in normal incidence) as $p_r \approx 2 \left< e_{ac} \right>$ where $e_{ac}$  is the acoustic energy density of the incident acoustic wave in the liquid\cite{Borgnis1953} and the brackets $<>$ correspond to time averaging. In the plane wave approximation, the acoustic energy is equally distributed between potential and kinetic energy and in space, so $\left< e_{ac}\right>=2 \left<e_c\right>$ with $e_c$ the kinetic energy density. Then, considering a harmonic acoustic wave, we have $\left< e_c \right> = 1/2 \rho_l V_l^2$ where $V_l$ is the amplitude of the acoustic velocity perturbation in the liquid and $\rho_l$ the liquid density. Finally, since the acoustic wave is radiated in the liquid with the Rayleigh angle $\theta_R$ , continuity of the velocity field at the substrate/drop interface gives: $V_l = A_s \omega_{ac} / cos(\theta_R)$, where $A_s$ is the amplitude of the normal acoustically induced displacement at the surface of the substrate, and $\omega_{ac}$ the frequency of the acoustic wave. Finally, the Laplace law gives the order of magnitude of the pressure drop at the interface due to capillary effects: $p_{cap} \approx 2 \sigma / R$ with $\sigma$ the liquid surface tension. As a consequence, the acoustical Weber number can be estimated according to the formula: $We_{ac} = p_r / p_{cap} = \rho_l A_s^2 \omega_{ac}^2 R / \sigma \cos^2(\theta_R)$. In this study we explored $We_\text{ac}$ ranging from 0.2 to 0.6.


\section{\label{sec:cfspd}Comparison of the dynamics of sessile and pendant drops: effect of gravity}

The observation of the dynamics of sessile droplet excited by traveling surface acoustic waves (see Movie 1 in SI) might give the misleading impression that during an oscillation cycle, the droplet is stretched by the effect of the wave (radiation pressure) and then falls down due to gravity. A previous study by Brunet et al. \cite{pre_brunet_2010} has shown that $f$ depends on the drop volume $V$ with a power law $f \sim V^{-1/2}$, which is typical for Rayleigh-Lamb inertio-capillary vibration modes \cite{prsl_rayleigh_1879,cup_lamb_1932} (a gravity-based restoring force would have led to a $V^{-1/6}$ power law,  see e.g. \cite{epl_perez_1999}). We can conclude from these results that capillary effects are dominant over gravity ones (as expected at such moderate Bo).  Nevertheless if gravity effects were negligible, dynamics of sessile and pendant drops should be strictly identical. We will show in the following that while it would be the case if the droplet was a linear oscillator, gravity strongly affects the drop dynamics due to nonlinearities. More generally, we will show that independently of its origin (external field, dynamically induced), the average drop stretching dramatically modifies the drop response. Before exploring this subject, we will first identify some specific regimes where the comparison between sessile and pendant drop dynamics is not relevant. Then we will compare the dynamics of oscillation and translation of sessile and pendant drops.

\subsection{\label{ss:spread}Specific regimes for pendant drops}

When pendant drops are stretched above a critical threshold during their oscillation cycle, they detach from the surface. Interestingly some new regimes of oscillation and displacement appear close to this threshold. We will first identify and analyze these regimes.

\subsubsection{Phase diagram \label{ss:pd}}

\begin{figure}[htbp]
\centerline{\includegraphics[width=0.45 \textwidth]{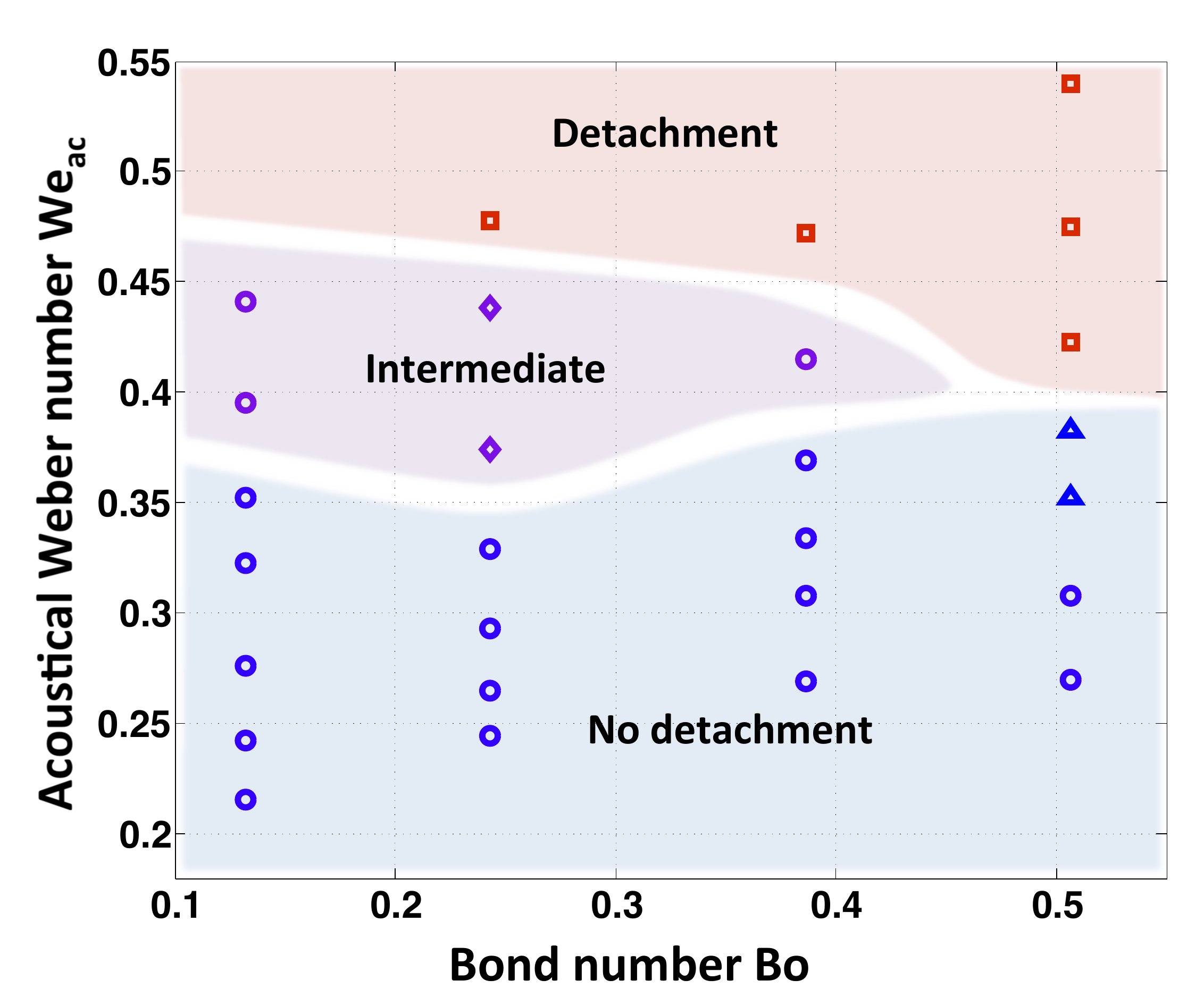}}
\caption{\label{fig:phased} Phase diagram as a function of $We_{ac}$ and $Bo$ summarizing the different regimes observed experimentally when pendant drops are excited by SAW. Blue region: no droplet detachment from the surface. Purple region: intermediate region with either detachment or sticking of the drop to the surface. Red region: systematic drop detachment.}
\end{figure}

\begin{figure}[htbp]
\centerline{\includegraphics[width=0.50 \textwidth]{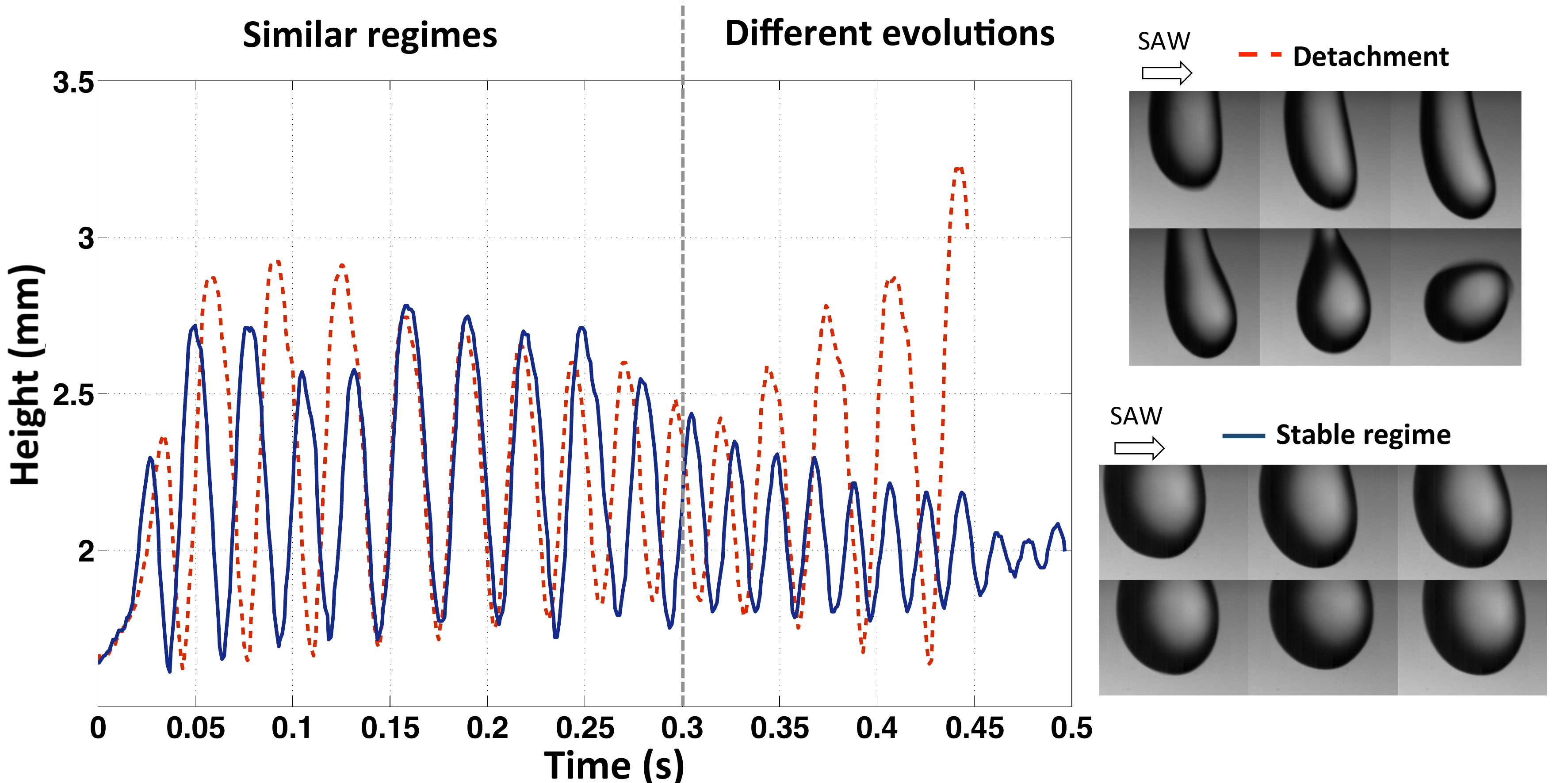}}
\caption{\label{fig:bifurcation} Evolution of the drop height for two droplets of the same size ($Bo = 0.24$) excited with the same acoustical signal ($We_\text{ac} = 0.44$). The drop can either detach from the substrate (red line) either reach a stable regime with small oscillations and larger translation speed.}
\end{figure}

The different droplet dynamics observed for pendant drops are summarized in the phase diagram of Fig. \ref{fig:phased}. For \textit{low $We_\text{ac}$} (blue region), the droplet always remains attached to the surface and undergoes coupled oscillatory and translational motions similar to the one observed for sessile droplet at the same acoustic power. In the \textit{intermediate region} (purple), small variations in the experimental conditions can either lead to the detachment of the drop or to its sticking to the walls. This high sensitivity to experimental initial conditions is especially evident for droplets of 5 $\mu$l (corresponding to $Bo = 0.24$):  for the same control parameters and after a similar transient regime, the droplet either experiences large oscillations leading to its detachment from the surface or small oscillations with larger translation speed (see Fig. \ref{fig:bifurcation}). Finally for \textit{high $We_\text{ac}$} detachment always occurs.

\begin{figure}[htbp]
\centerline{\includegraphics[width=0.50 \textwidth]{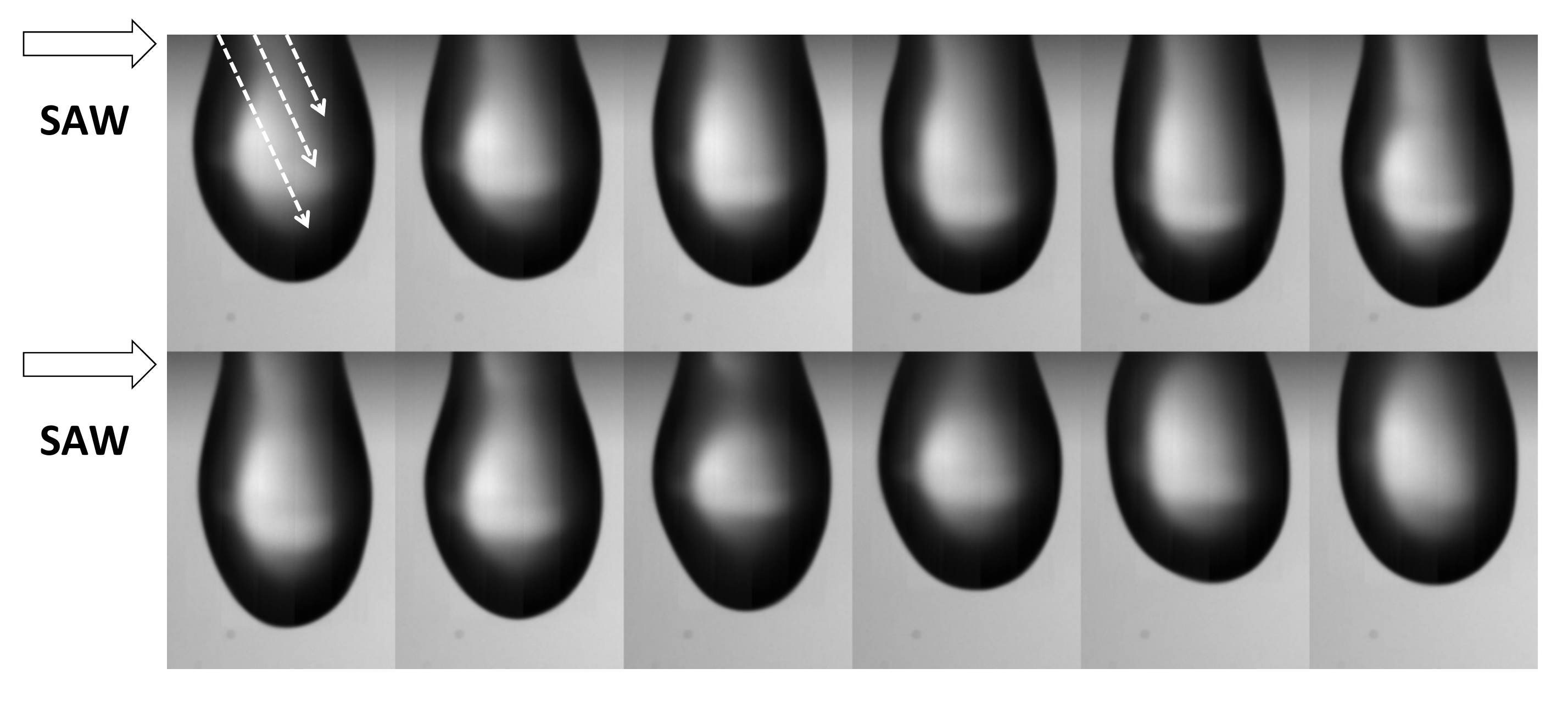}}
\caption{\label{fig:stable} Sequence of images (time interval 4 ms) of a drop of volume $15 \mu l$ ($Bo = 0.5$) excited with an acoustic power corresponding to $We_{ac} = 0.38$. The drop reaches a stable regime with large deformation but almost no oscillation and translation.}
\end{figure}

It is interesting to note that a peculiar regime is observed for the biggest drop (15 $\mu$l, $Bo = 0.5$) close to the detachment threshold (triangles on Fig. \ref{fig:phased}). In this regime, the droplet reaches an equilibrium state with extremely small oscillations and slow velocity. The drop shape is quasi-axisymmetric (explaining why there is no translation) and the drop stretching remains close to the detachment threshold (see Fig. \ref{fig:stable} and Movie 3 in SI).

\subsubsection{\label{ss:d}Detachment}

\begin{figure}[htbp]
\centerline{\includegraphics[width=0.45 \textwidth]{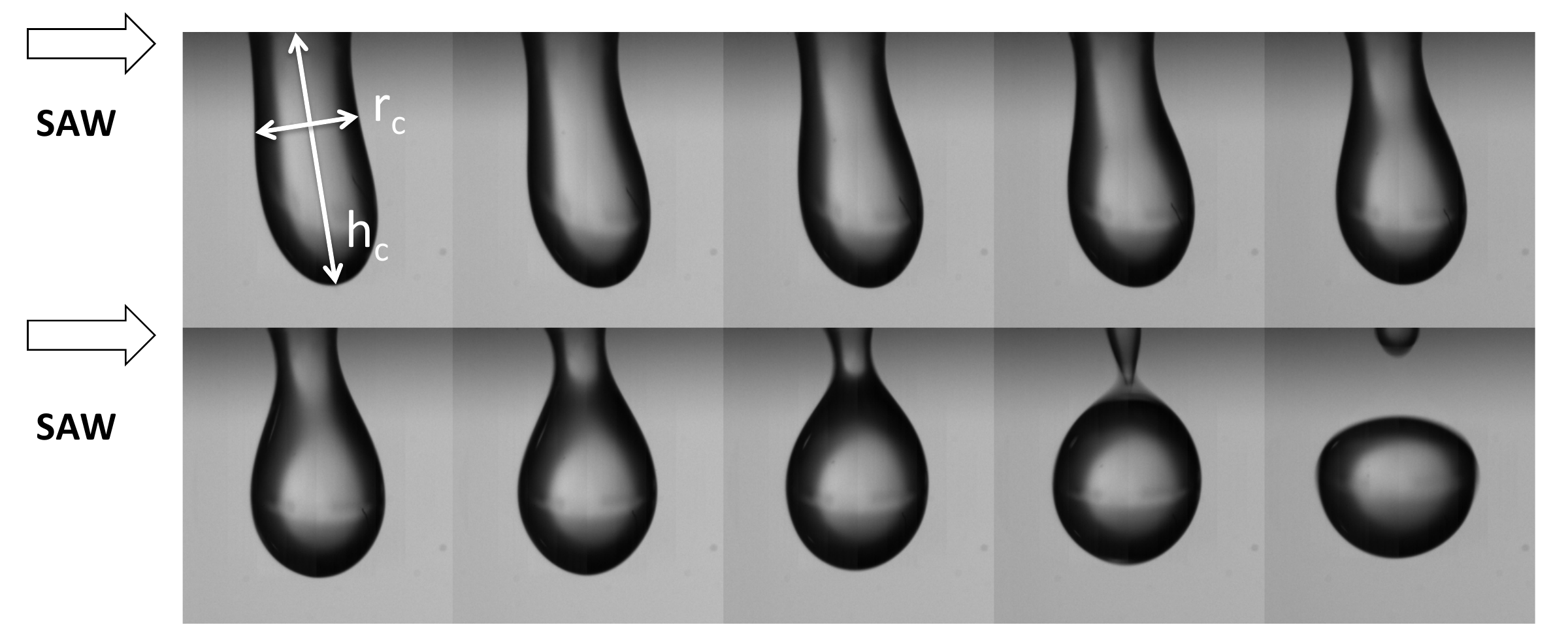}}
\caption{\label{fig:detachment} Sequence of images (time interval 2 ms) showing the detachment of a drop of $10 \mu l$ excited with an acoustic power corresponding to $We_{ac} = 0.41$.}
\end{figure}

Droplet detachment from the substrate is illustrated on Fig. \ref{fig:detachment}. At first the drop is stretched vertically forming a quasi-cylindrical liquid column. Then the base of the drop is squeezed and pinches off. It is well known since the work of Savart \cite{acp_savart_1833}, Plateau \cite{mars_plateau_1849} and Rayleigh \cite{plms_rayleigh_1878,prsl_rayleigh_1879}, that liquid columns are unstable to the Rayleigh-Plateau instability. Savart was the first to observe the decay of liquid jet into drops. Then Plateau has shown that surface tension favors the development of long wavelength undulations at the surface of a liquid column since they reduce its free surface energy. Nevertheless an analysis based solely on surface tension would predict the formation of the largest droplets (since they minimize the surface energy) while this is not observed in practice. Later on, Rayleigh demonstrated that surface tension has to work against inertia to induce jet breakup. With a stability analysis, he proved that the most unstable perturbations corresponds to wavelength $\lambda_R \approx 9 R_o$ and that the characteristic time associated with this instability is $T_R = 2.91 \sqrt{\frac{\rho_l R_o^3}{\sigma}}$ with $R_o$ the liquid jet radius, $\sigma$ the surface tension and $\rho_l$ the liquid density. In the present experiments, we measured the critical aspect ratio $h_c/R_c$ of the liquid column during the last oscillation before breakup occurs ($h_c$ and $R_c$ are respectively the critical height and radius of the liquid column before breakup occurs). 

\begin{figure}[htbp]
\centerline{\includegraphics[width=0.40 \textwidth]{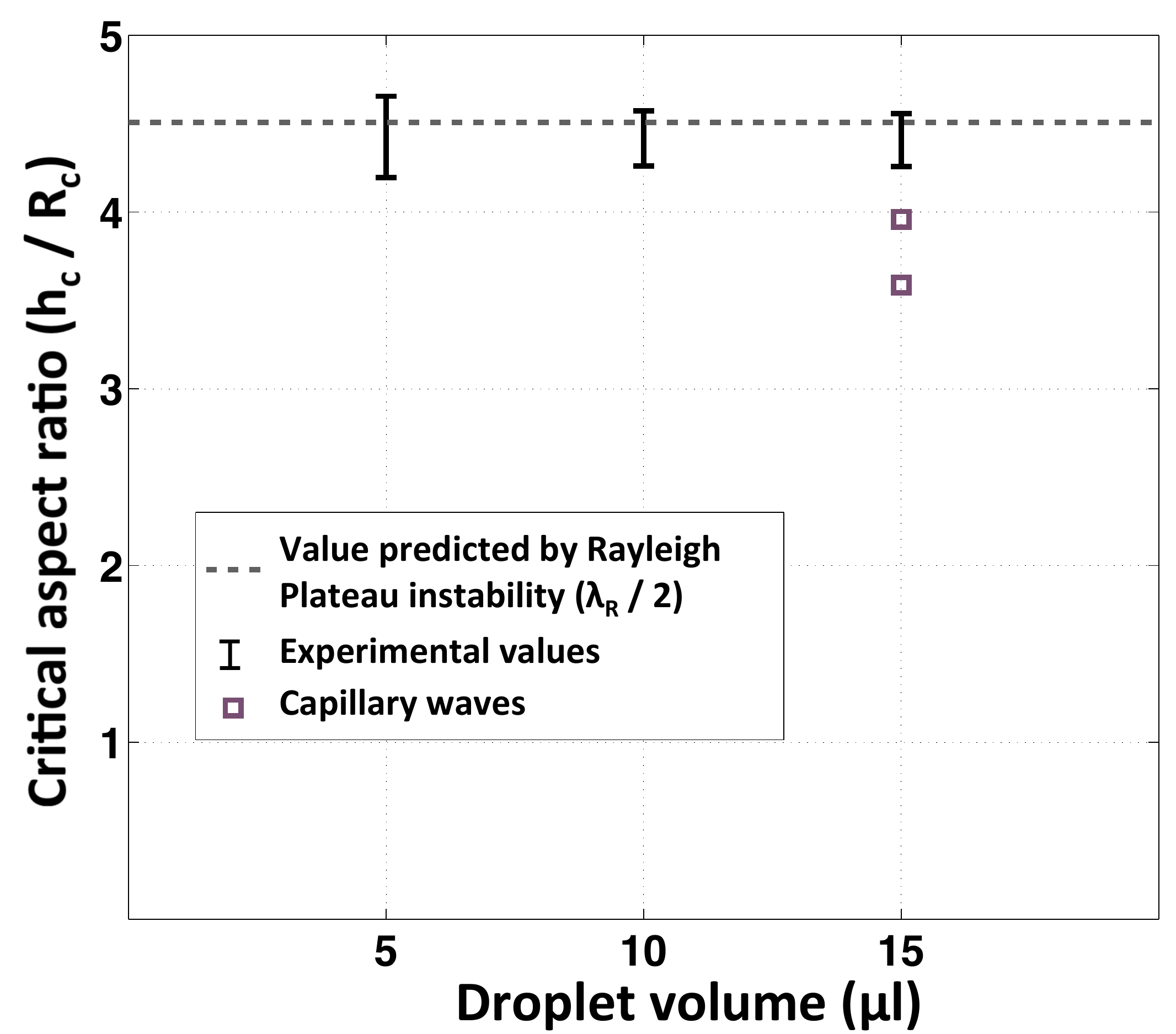}}
\caption{\label{fig:rayleighplateau} Critical aspect ratio $h_c / R_c$ leading to the drop detachment from the surface for different initial volumes. Square symbols correspond to the peculiar regime shown in Movie 4 in SI.}
\end{figure}

Independently of the initial drop volume, we found that droplet detachment occurs when $h_c/ R_c \approx 4.5$ (i.e. $\lambda_R /2$) (see Fig. \ref{fig:rayleighplateau}). The same tendency was observed in ref. \onlinecite{prl_tan_2009} for focused SAWs. An interesting point is that the development of Rayleigh-Plateau instability requires not only a sufficient drop elongation but also a long enough time for the instability to develop. But since drop oscillations are inertio-capillarity oscillations, their period also scale as $\sqrt{\frac{\rho_l R_o^3}{\sigma}}$, which means that the two phenomena (instability and drop oscillations) have comparable characteristic times in the invicid regime.

As we can see in figure \ref{fig:rayleighplateau}, two detachments deviate from the value predicted by the theory. Those points correspond to the peculiar regime presented in figure \ref{fig:stable} but for a slightly higher $We_\text{ac}$. In this case, droplets experience stable oscillation during a long time and finally detach in a quasi-static way (see Movie 4 in SI).

\subsection{\label{ss:dom}Droplet oscillatory motion}

Now we can compare the dynamics of sessile and pendant drops away from these specific regimes. In this case, droplets remain attached to the substrate and undergo both oscillations and a translational motion. In this subsection we will focus on the effect of gravity on droplet quadrioplar oscillations. In the next subsection we will investigate the droplet translational motion. 

\begin{figure}[htbp]
\centerline{\includegraphics[width=0.45 \textwidth]{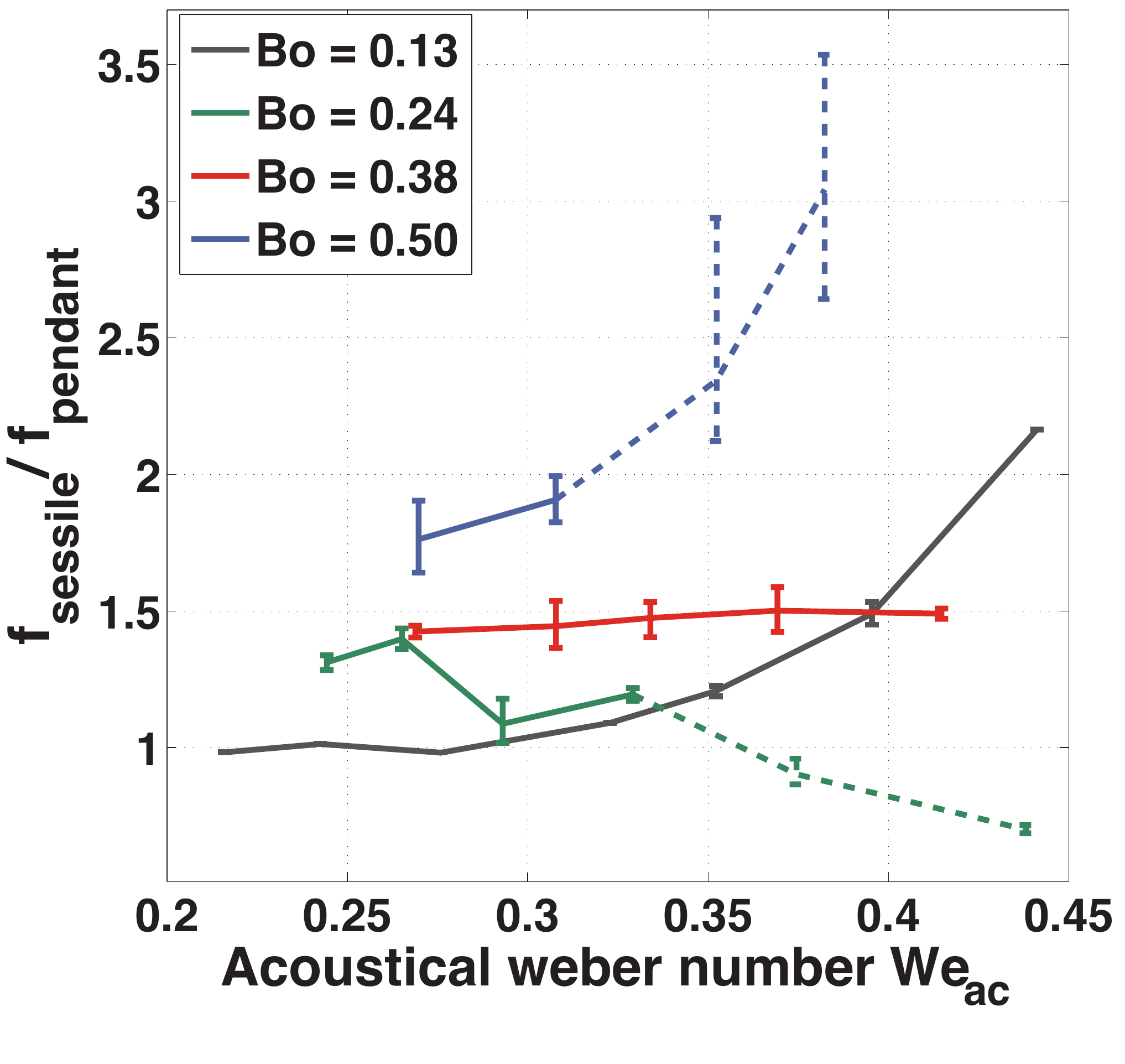}}
\caption{\label{fig:freqcomp} Ratio between sessile and pendant drops oscillation frequencies $f_{\mbox{sessile}} / f_{\mbox{pendant}}$ as a function of $We_{ac}$ for different $Bo$ (corresponding to different droplet sizes). The specific regimes appearing close to the droplet detachment threshold are represented with dashed lines.}
\end{figure}

A direct comparaison of the quadrupolar oscillations for sessile and pendant drops at the same driving parameters shows that $f_{\text{osc}}$ is always higher in the first case than in the second one (see Fig. \ref{fig:freqcomp}). This discrepancy increases with Bo. Nevertheless, even at small initial Bond number, gravity can dramatically affect the oscillatory dynamics of the drop (factor up to 2 on $f_{\text{osc}}$) when it experiences large deformations (at large $We_\text{ac}$).

\subsubsection{Theoretical analysis}

To understand this behavior, we will consider that droplet vibrations can be described by an oscillator equation. This has been demonstrated for the eigen modes of levitating droplets \cite{Trinh1982,Tsamopoulos1983,Bussonniere2014} and sessile droplets with pinned contact lines \cite{Bostwick2014,Chang2015}. The extension of such equations to sessile droplet with moving contact lines would require to properly include the dissipation in the vicinity of the triple line \cite{Lyubimov2006a}. For small oscillations of the drop, the associated oscillator equation is linear and thus gravity does not affect its eigenfrequency but only its equilibrium position. Indeed, if we consider a harmonic oscillator submitted to the effect of gravity: $m \ddot{x} = - k x + mg$ (with $k$ the spring stiffness, $m$ the mass of the system and $g$ the standard gravity) and introduce the natural angular frequency of the system $\omega_o = \sqrt{\frac{k}{m}}$, a simple change of variables $X = x-x_s$ with $x_s = g / \omega_o^2$ the static equilibrium position gives:
$$
\ddot{X} + \omega_o^2 X = 0,
$$
showing that the natural frequency $\omega_o$ is not affected by the external constant force field $mg$. Nevertheless it has been proved \cite{Tsamopoulos1983,Bussonniere2014,Lyubimov2006a} that droplets are nonlinear oscillators with quadratic nonlinearities when they undergo finite amplitude oscillations. These nonlinearities are due to the convective nonlinearity of Navier-Stokes equations but also to the nonlinearities associated with the interface deformation. 

For the sake of simplicity and to qualitatively illustrate the effect, we will first consider an oscillator with a simple quadratic nonlinearity $\alpha x^2$ submitted to a constant force field $mg$: $\ddot{x} + \omega_o^2 (1 + \alpha  x)x = g$ with $\alpha$ the nonlinearity coefficient. If we introduce as previously the static equilibrium position $x_s$, solution of the equation $x_s + \alpha x_s^2 = g / \omega_o^2$, and make the substitution: $X = x-x_s$, we get:
\begin{equation}
\ddot{X} + \omega_o^2 (1 + 2 \alpha x_s + \alpha X)  X = 0.
\label{eq:osc1}
\end{equation}
In this case, we clearly see that both the static external force field $mg$ (leading to a shift of the equilibrium position $x_s$) and dynamical effects affect the eigenfrequency of the system. We will analyze these two effects separately. If we consider tiny oscillations of the system $X \ll x_s$, around $x_s$, the eigen frequency of the system $\omega_s$ becomes: $\omega_s^2 = \omega_o^2 (1 + 2 \alpha x_s)$, that is to say for moderate droplet deformation $\alpha x_s \ll 1$:
$$
\omega_s = \omega_o (1 + \alpha x_s).
$$
From a physical point of view, the effect of the external force field is clear: since the string is nonlinear and depends on the oscillator stretching, the gravity field simply modifies the equilibrium position of the system and thus its eigenfrequency.

To understand the role of dynamical effects, we can consider the situation when $x_s \ll X$. In this case, Eq. (\ref{eq:osc1}) becomes simply the one of a nonlinear oscillator with quadratic nonlinearities:
$$
\ddot{X} + \omega_o^2 (1 + \alpha X)  X = 0.
$$
This problem is treated in Landau textbook \cite{Landau1976}. Poincar\'{e} expansion of $X$: $X = \epsilon X_1 + \epsilon^2 X_2 + \epsilon^3 X_3$, with $X_1 = A \cos (\omega_d)$ and $\omega_d = \omega_o + \epsilon \omega_1 + \epsilon^2 \omega_2$ gives:
$$X_2 = - \frac{\alpha A^2}{2} + \frac{\alpha A^2}{6} \cos(2 \omega_d t).$$
We obtain the classical result that a quadratic nonlinearities leads to a static deformation $x_{d} = - \alpha A^2 / 2$ and oscillations at frequency $2 \omega_o$. With this asymptotic expansion, we can also compute the eigen frequency shift induced by nonlinear effects: $\omega_d = \omega_o ( 1 + 5/6 \, \alpha x_d)$. 

Now, if we combine the effects of the stationary force field and dynamical effects we simply obtain: $\omega_{nl} = \omega_o \left(1 + \alpha (x_g + 5/6 x_d) \right)$, where $\omega_{nl}$ is the eigen frequency of the nonlinear oscillator. It is interesting to note that both effects depend on the average stretching of the spring-mass system. Nevertheless, while the frequency shift induced by a steady external force field is independent of the amplitude of oscillation of the system, the frequency shift induced by dynamical effects is proportional to $A^2$ (since $x_d \propto A^2$).

The same method applies to the nonlinear equation describing drop quadrupolar oscillations \cite{Bussonniere2014}: 
\begin{equation}
\ddot{x} + 2 \lambda \dot{x} + \omega_o^2 x + \alpha \omega_o ^2 x^2 + \beta \dot{x}^2 + \gamma x \ddot{x} = 0.
\label{NLequa}
\end{equation}
In this case, proper analysis (see appendix) shows that:
\begin{eqnarray}
\omega_{nl} & = & \omega_o \left( 1 + (\alpha - \gamma/2) x_s + K x_d \right),
\end{eqnarray}
where $K$ is a complex fonction of the nonlinear coefficients $\alpha$, $\beta$ and $\gamma$, and $x_d \propto A^2$. Of course, this analysis holds for weakly nonlinear systems. For larger nonlinearities, the dependance of the eigen frequency over the average drop stretching might be more complex than a simple linear shift.

\subsubsection{Experimental results}

\begin{figure}[htbp]
\centerline{\includegraphics[width=0.5 \textwidth]{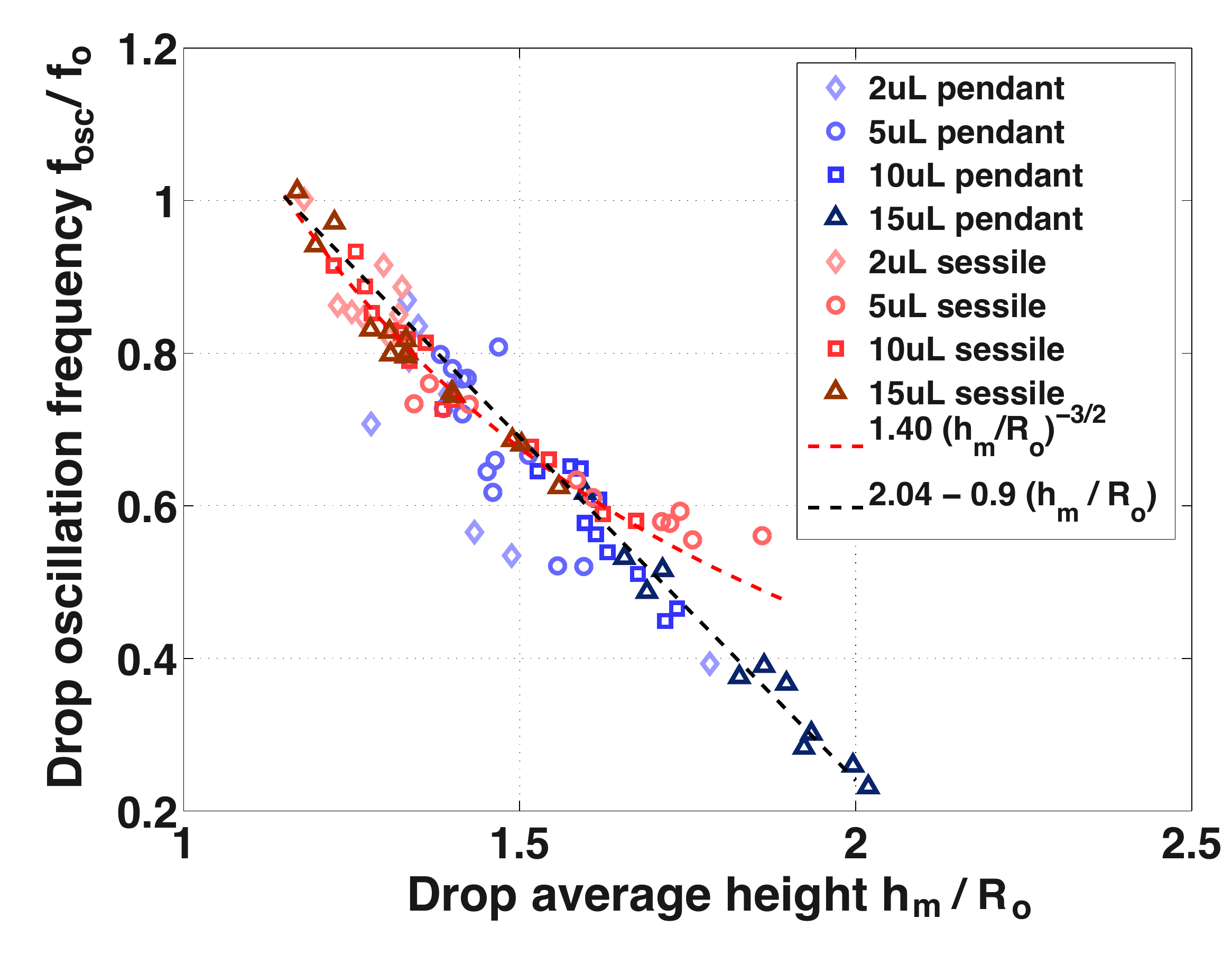}}
\caption{\label{fig:fehm} Eigen frequency $f_{\text{osc}}$ divided by the Rayleigh-Lamb frequency $f_o$ as a function of the drop average height $h_m$ divided by the drop static radius $R_o=(3 V / 2 \pi)^{1/3}$ in absence of gravity effects for different drop volumes and acoustical powers.}
\end{figure}

The dependance of the eigenfrequency $f_{\text{osc}}$ on its average stretching $h_m = 1/T \int_0^T h(t) dt$  has been measured for different acoustical power and droplet sizes (see Fig. \ref{fig:fehm}). When $f_{\text{osc}}$ divided by $f_o = 1 / 2 \pi \sqrt{8 \sigma / \rho_l R_o^3}$ (predicted by Rayleigh-Lamb theory) is plotted as a function of $h_m$ divided by $R_o = (3 V / 2 \pi)^{1/3}$ (the drop radius in absence of gravity effects), all data collapse into a single curve for both sessile and pendant drops as expected from previous analysis \footnote{NB: Since the contact angle is slightly superior to $\pi / 2$ and the Rayleigh-Lamb frequency is estimated for a levitating drop while it is sessile or pendant here, we do not have exactly $f/f_o \rightarrow 1$ when $h_m / R_o \rightarrow 1$.}.  Nevertheless, the power law differs in these two cases. Data points are well fitted by a linear law for pendant drops $(f_{\text{osc}} / f_o \approx 2.04 - 0.9 h_m/R_o)$ and by a power law $f_{\text{osc}} / f_o = 1.4 (h_m / R_o)^{-3/2}$ for sessile ones. 

\begin{figure}[htbp]
\centerline{\includegraphics[width=0.5 \textwidth]{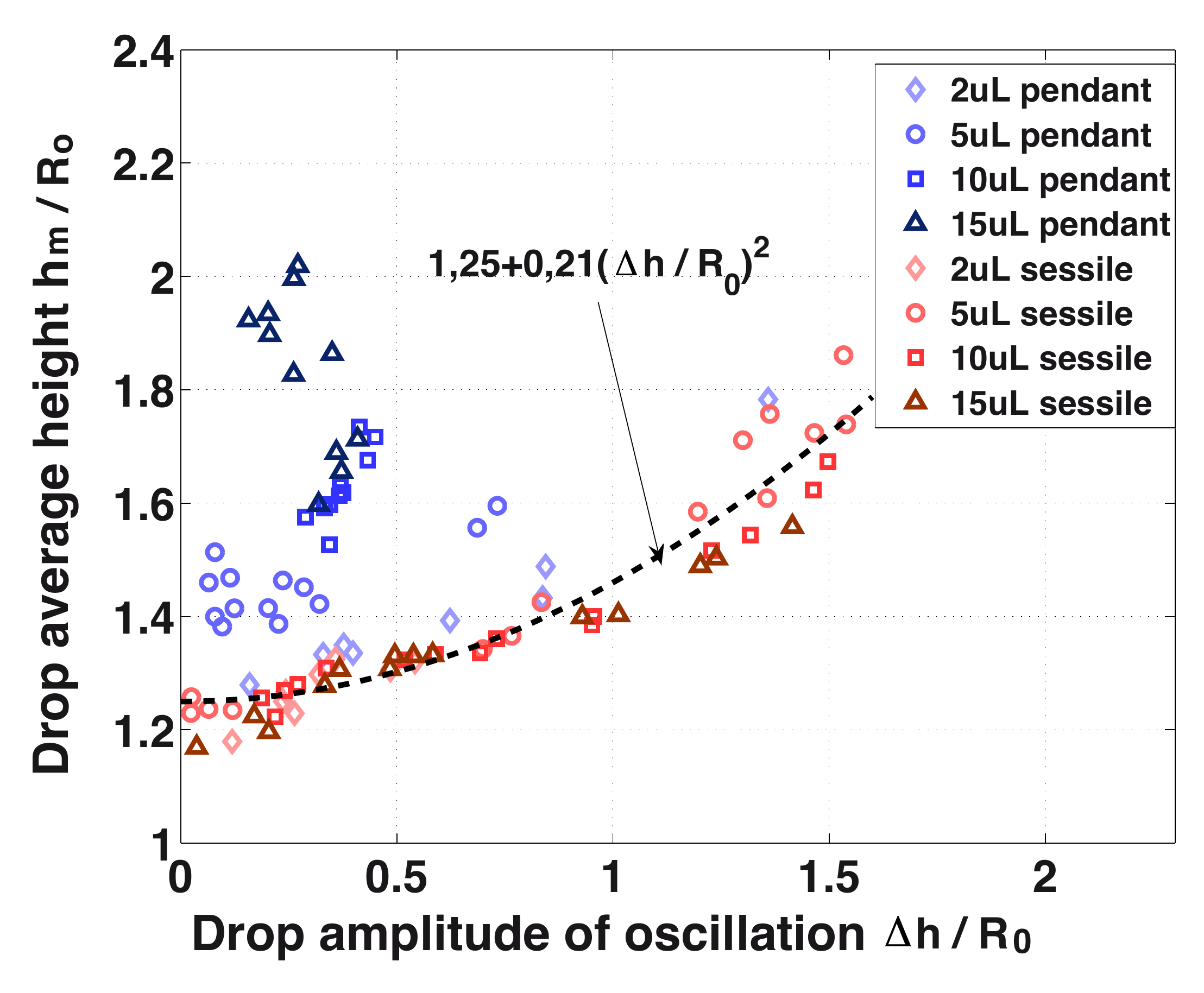}}
\caption{\label{fig:hmdh} Drop average height $h_m$ as a function of the drop amplitude of oscillation $\Delta h$  both divided by $R_o=(3 V / 2 \pi)^{1/3}$, the drop static radius in absence of gravity. Only data for sessile drops show clear correlation between both quantities.}
\end{figure}

To determine which effects are responsible for the drop stretching (external force field or dynamical effects), we plotted the dimensionless average droplet stretching $h_m / R_o$ as a function of the droplet amplitude of oscillation $\Delta h / R_o$ (see Fig. \ref{fig:hmdh}). For pendant drops no correlation between these two parameters is observed (except for the smallest drops of 2 $\mu L$) while for sessile droplets, a power law $h_m / R_o \approx 1.25 + 0.21 (\Delta h / R_o)^2$ is obtained as expected from previous analysis when dynamical effects are dominant. Thus, these experiments show that the average drop stretching is mainly caused by stationary external force fields in the case of pendant drop, whereas dynamical effects induced by large amplitude oscillations are dominant for sessile drops. Indeed, in the first case, gravity and the radiation pressure induced by the acoustic field act in the same direction and lead to large $h_m$. For sessile drops however, they act in opposite directions, and the average deformation induced by the drop oscillations and nonlinearities becomes dominant. Of course, for the smallest $Bo$ (tiniest drops) the same evolution is observed for sessile and pendant drops since in this case gravity forces are negligible compared to capillary effects.

With these experimental results we can now explain the trends observed on Fig. \ref{fig:freqcomp}. A higher frequency for sessile drops at constant $We_\text{ac}$ is simply the result of smaller averaged deformations compared to those of pendant drops, which lead to a smaller frequency shift induced by nonlinear effects. 

In conclusion, gravity acts indirectly on the drop oscillation frequency through the nonlinearities of the drop vibrations. The frequency shift can be significant (factor up to 2.5) even at moderate Bond numbers. Moreover, inverting the gravity field does not only affect the value of the eigen frequency, but also its dependance over the amplitude of the drop oscillations. 

\begin{figure}[htbp]
\centerline{\includegraphics[width=0.5 \textwidth]{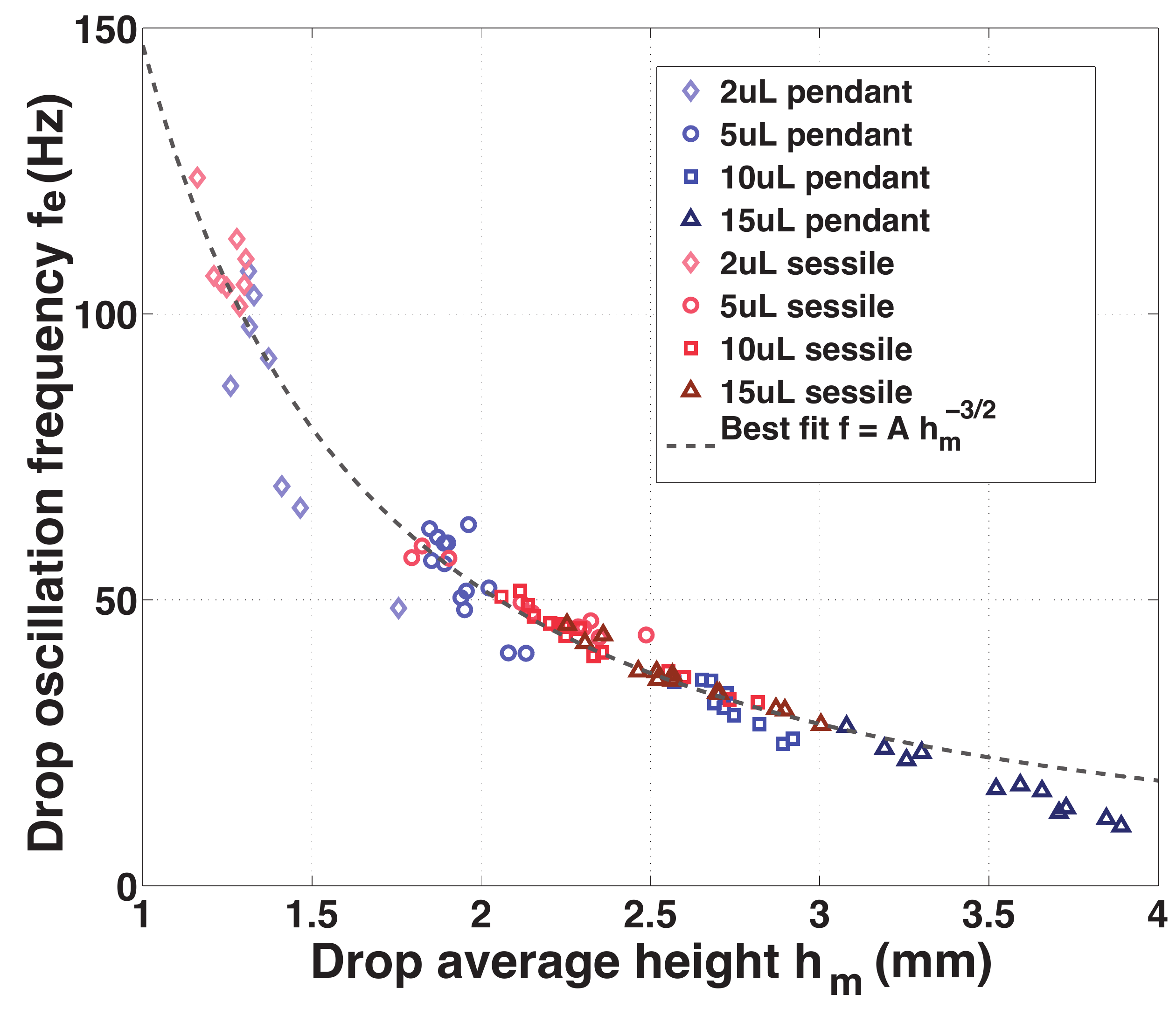}}
\caption{\label{fig:fmhm} Drop oscillation frequency as a function of the drop average height $h_m$ for sessile (red) and pendant (blue) drop of different volumes, and different driving acoustical power. All data collapse into a single trend close to the power law predicted by the Rayleigh-Lamb theory.}
\end{figure}

Finally, an interesting result is that it is possible to collapse all data obtained for both pendant and sessile droplets, different acoustical powers and droplet sizes by simply replacing the radius by the average height $h_m$ in Rayleigh-Lamb formula (see Fig. \ref{fig:fmhm}). It simply means that the characteristic length for the computation of stretched droplets eigen frequency is no longer the drop radius but instead their average height.

\subsection{\label{ss:dtm}Droplet translational motion and correlation with the oscillations}


When a droplet is excited by traveling surface acoustic waves, it moves in the same direction as the  wave. This translation is due to the asymmetry of the acoustic field radiated in the drop along Rayleigh angle (given by Snell-Descartes law) which induces asymmetric deformation of the drop and thus different contact angles at the front and rear parts of the drop contact line. 

\begin{figure}[htbp]
\centerline{\includegraphics[width=0.5 \textwidth]{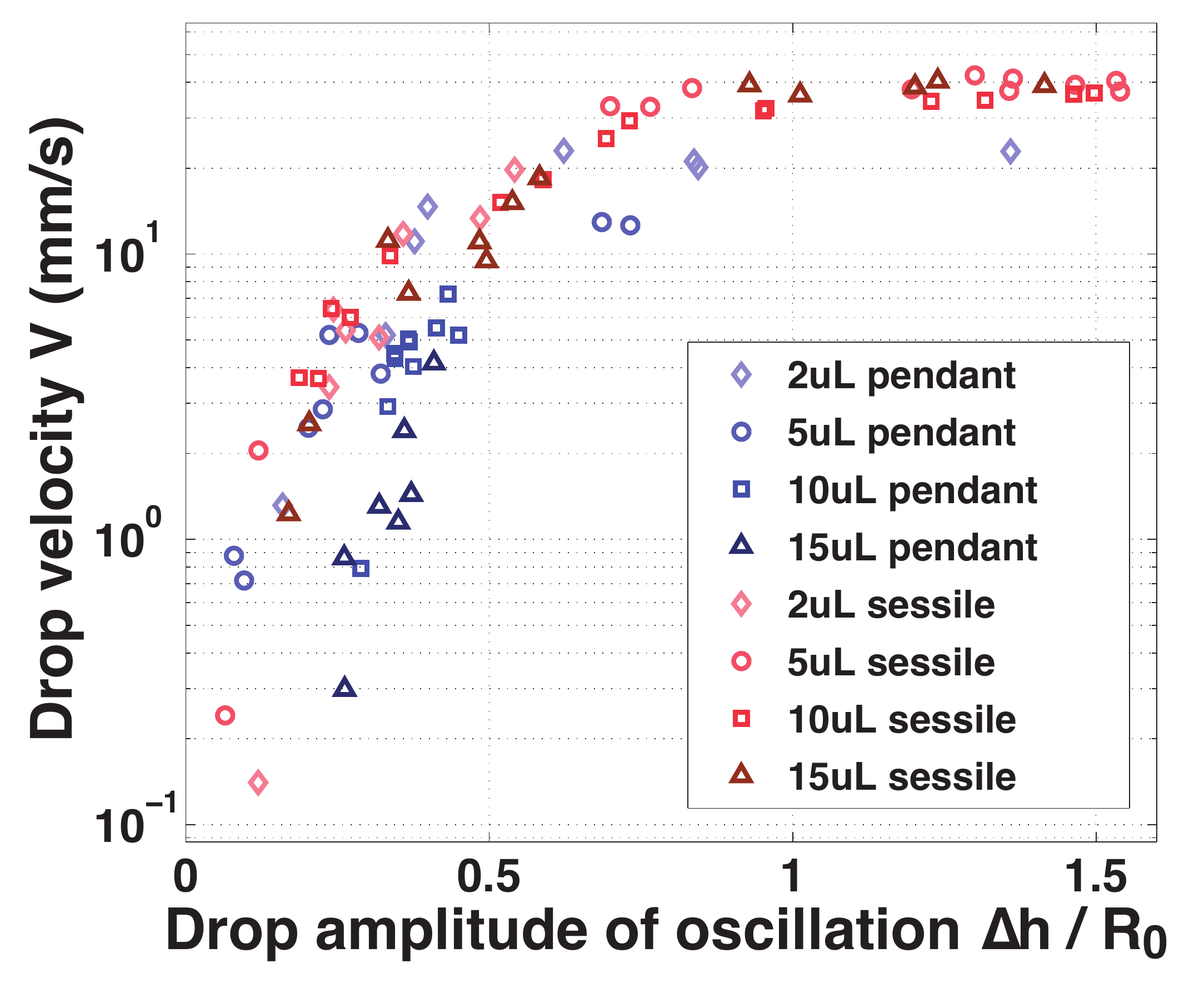}}
\caption{\label{fig:vdh} Sessile (red) and pendant (blue) drops translation speed as a function of the dimensionless amplitude $\Delta h / R_o$.}
\end{figure}

In a recent paper, we have shown that there is a strong correlation \cite{apl_baudoin_2012} between droplet translational and oscillatory motion of sessile droplet. This correlation is represented on Fig. \ref{fig:vdh} for sessile and pendant drops. The good collapse of the data for sessile droplet and the smallest pendant drops ($2 \; \mu L$ and $5 \; \mu L$) indicates that in these cases, the amplitude $\Delta h$ is the main quantity ruling its translational velocity. Whereas at higher volume ($10\ \mu L$ and $15\ \mu L$) pendant drop velocities $V$ deviate from this correlation. It is also interesting to note that the translational velocity increases rapidly for drop oscillation $\Delta h / R_o < 0.3$ and then a saturation occurs for larger oscillations. 


To understand this trend, we investigated the motion of the contact line during an oscillation cycle for both pendant and sessile drops. Indeed, $V$ can be seen as the product of $f$ times the net motion per period. 

\begin{figure}[htbp]
\centerline{\includegraphics[width=0.5 \textwidth]{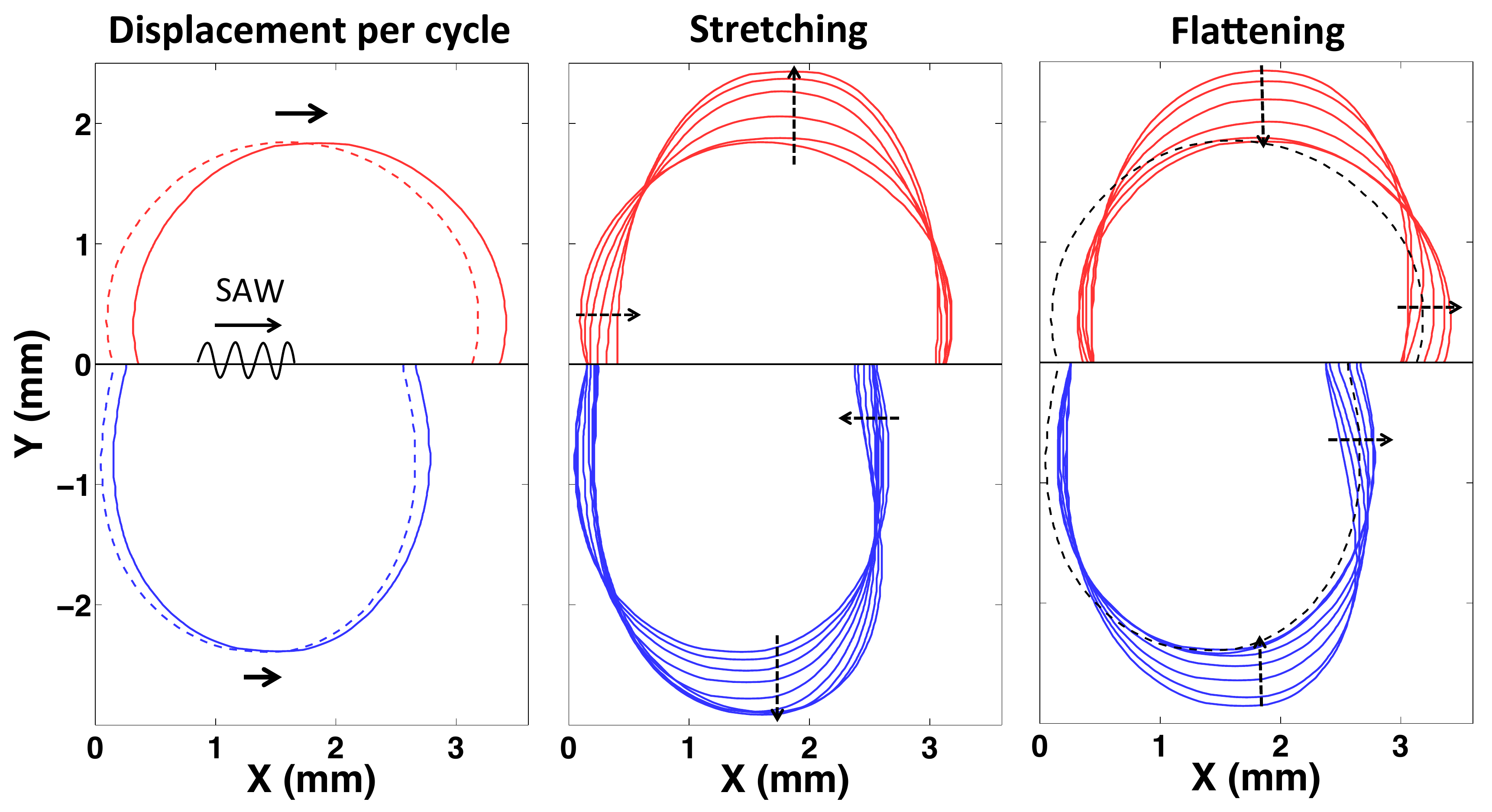}}
\caption{\label{fig:shape} Sessile (red) and pendant (blue) drop shape evolution during an oscillation cycle. Left: initial and final shape. Center: shape evolution during the drop stretching phase. Right: shape evolution during the flattening phase.}
\end{figure}

Figure \ref{fig:shape} compares the time-evolution shapes for sessile and pendant drops during a period at the same acoustic power, for the same drop size and for the same oscillation amplitude (see also Movie 5 in SI). For sessile drops, the contact line at the rear and that at the front move alternatively forward during the drop stretching and flattening (with much smaller displacement of the opposite contact line). However for pendant drops, forward displacement of the rear contact line essentially occurs during the drop elongation while the front contact line first moves backward and then forward. This picture is representative of most of the cycles observed for different acoustical powers and droplet sizes. 

\begin{figure}[htbp]
\centerline{\includegraphics[width=0.35 \textwidth]{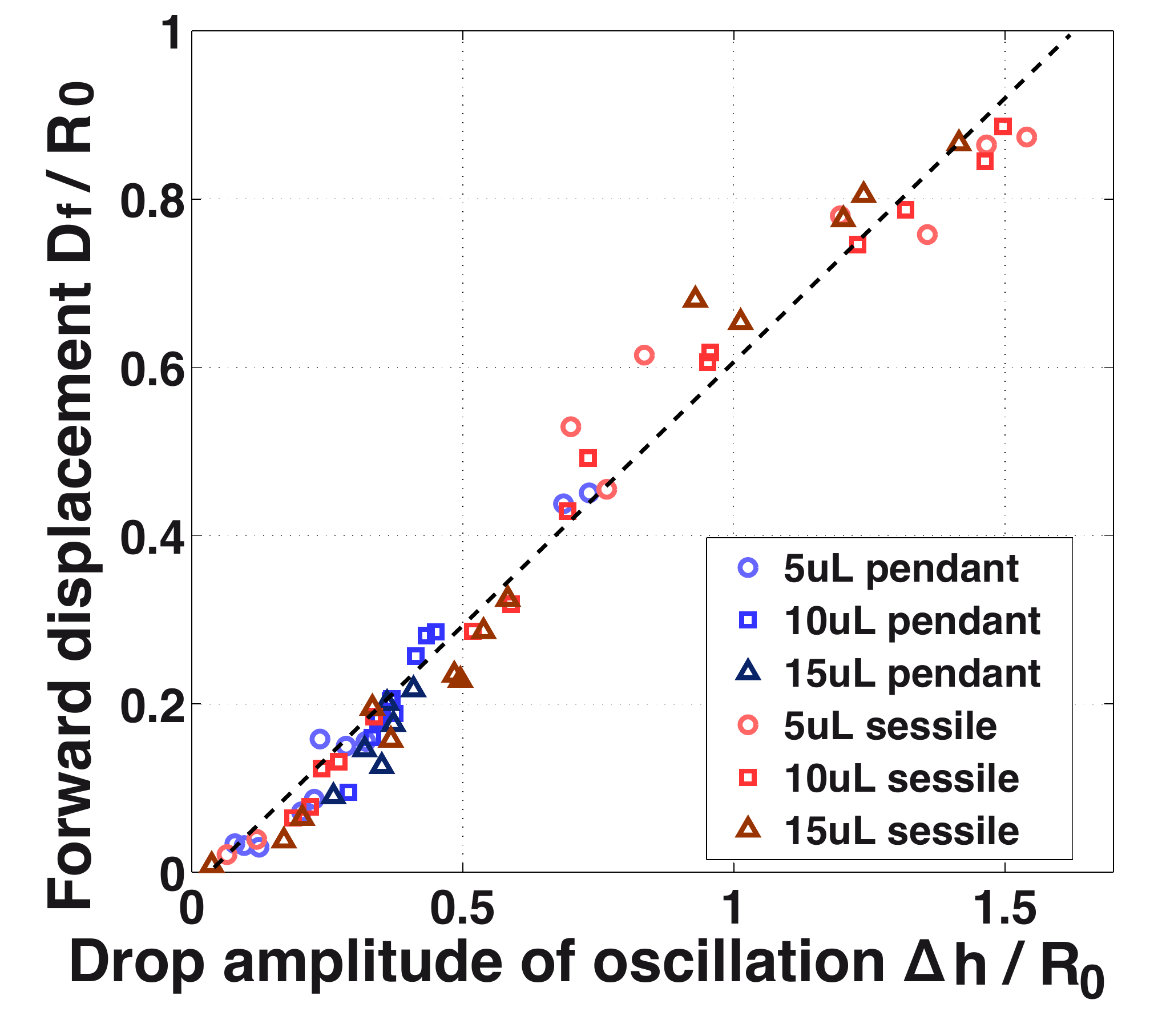}}
\centerline{\includegraphics[width=0.35 \textwidth]{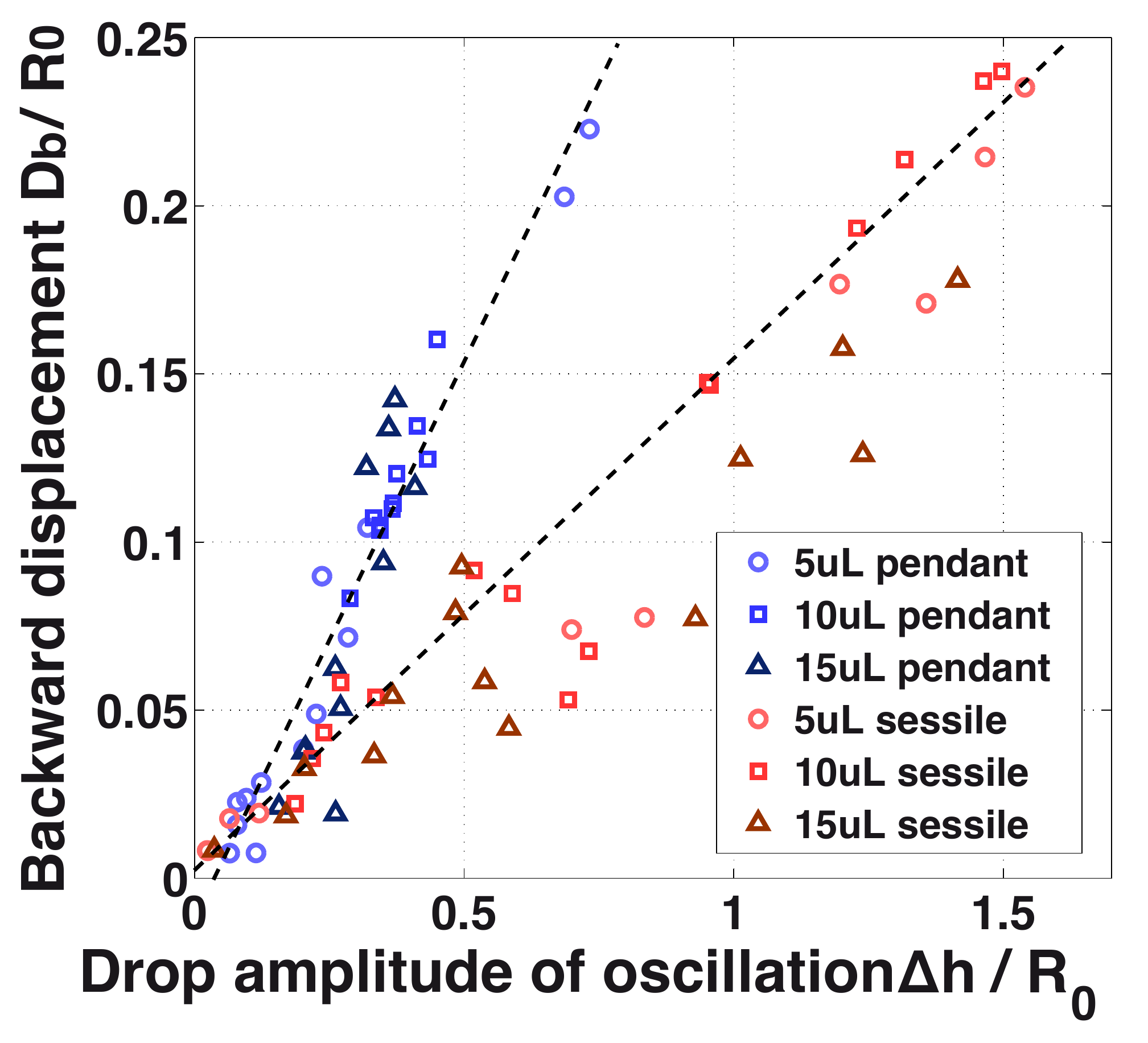}}
\caption{\label{fig:Ddh} Forward (up) and backward (down) displacement of the front part of the contact line of sessile (red) and pendant (blue) drops as a function of the dimensionless amplitude of oscillation $\Delta h / R_o$.}
\end{figure}

Then, we analyzed the relation between the net motion of the front contact line during the drop extension and flattening as a function of $\Delta h / R_o$ (see Fig. \ref{fig:Ddh}). The contact line displacement per period is essentially a linear function of the amplitude of oscillation during both the stretching (backward movement of the contact line) and the flattening (forward movement of the contact line). Sessile and pendant drops front contact lines displacement have the same dependency on the amplitude of oscillation during the stretching phase but as seen on Fig. \ref{fig:Ddh}, the backward motion is smaller at same amplitude of oscillation for sessile drops. Since the frequency of oscillation is also smaller for pendant drops at the same amplitude of oscillation, it explains why pendant droplet velocity is always smaller as observed on Fig. \ref{fig:vdh}.

\section{Semi-empirical law and origin of the saturation for sessile drops}

\begin{figure}[htbp]
\centerline{\includegraphics[width=0.45 \textwidth]{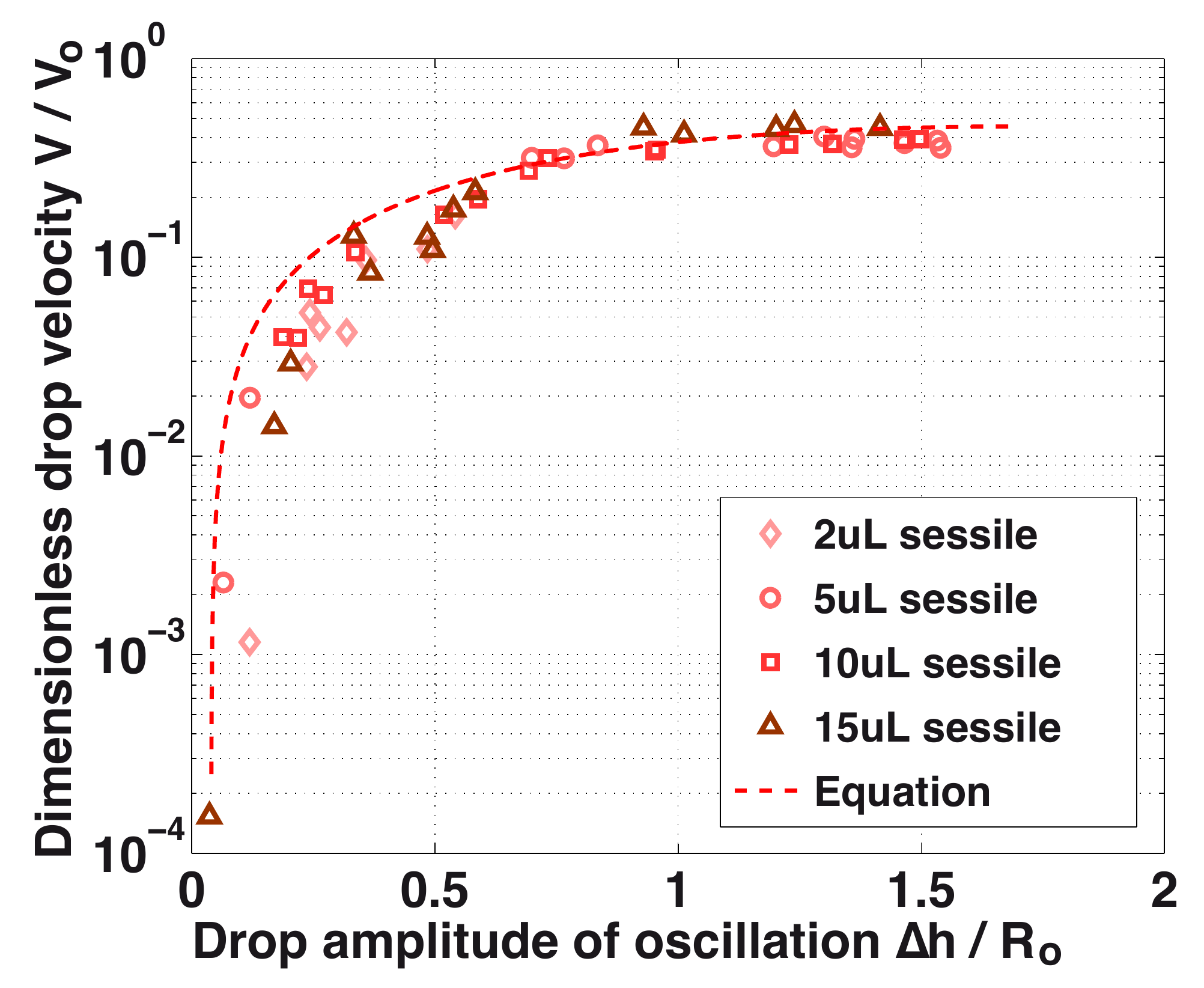}}
\caption{\label{fig:vdhf} Dimensionless sessile droplet translation speed $V_/ V_o$ as a function of the drop amplitude of oscillation $\Delta h / R_o$ with $V_o = f_o R_o$ the characteristic velocity associated with droplet oscillations and $R_o=(3 V / 2 \pi)^{3/2}$, the drop static radius in absence of gravity. Dashed line: equation (\ref{eq:final}).}
\end{figure}

In the previous sections, we have shown successively that:
\begin{itemize}
\item The drop translation speed can be seen as the product of the displacement per cycle $D$ times the frequency of oscillation: $V = D \times f_{\text{osc}}$,
\item The drop displacement per cycle is essentially a linear function of the drop amplitude of oscillation: $D = K_1 \times \Delta h / R_o + K_2$,
\item The oscillation frequency of sessile drops is related to the mean drop height according to a formula: $f/f_o = K_3 (h_m / R_o)^{-3/2}$, which itself depends on the square of the drop amplitude of oscillation: $h_m/R_o = K_4 + K_5 (\Delta h / R_o)^2$,
\end{itemize}
with the coefficients $K_1 = 0.50$, $K_2 = 0.02$, $K_3 = 1.41$, $K_4 = 1.25$ and $K_5 = 0.21$ determined experimentally. If we combine these relations, we obtain:
\begin{equation}
 \frac{V}{V_o} = C_1 \left[ \frac{\Delta h}{R_o} + C_2 \right] \times \left[ 1 + C_3 \left( \frac{\Delta h}{R_o} \right)^2 \right]^{-3/2},
\label{eq:final}
\end{equation}
with $V_o = f_o R_o$ the characteristic velocity associated with droplet oscillation, $C_1 = K_1 K_3 K_4^{-3/2} \approx 0.5$, $C_2 = K2/K_1 \approx - 0.04$ and $C_3 = K_5 / K_4 \approx 0.17$. 

The predictions obtained with eq.~(\ref{eq:final}) are compared to experimental data on Fig. \ref{fig:vdhf}. The trends are globally recovered even if a weak discrepancy is observed for intermediate amplitude oscillations. These differences comes from the fact that the net motion per period is not perfectly a linear function of $\Delta h$. Nevertheless this comparison allows us to understand that the saturation of the droplet velocity is due to the decrease of  $f$ induced by dynamical nonlinear effects.

\section{\label{sec:c}Conclusion}
In this paper, we have analyzed the effect of gravity on the dynamics of drops excited by SAWs but also clarified the link between droplets oscillations and their translational motion.  Further investigations using different contact angles and hysteresis would be of interest to get a deeper insight into the relation between drop oscillation and contact line mobility. Moreover, as underlined in a recent review \cite{arfm_yeo_2014}, another great challenge is to unveil the origin of the drop oscillations and establish the missing link between the high frequency acoustic excitation ($\sim 20$ MHz) and the low frequency ($\sim 50$ Hz) droplet oscillatory response. 

\section*{Acknowledgments}

This work was supported by grants from the Agence Nationale de la Recherche ANR-12-BS09-021-01, the Direction G\'{e}n\'{e}rale de l'Armement (France) and the R\'{e}gion Nord Pas-de-Calais

\appendix*
\section{}

Oscillating droplets can be modeled by the following nonlinear equation \cite{Bussonniere2014} :

\begin{equation}
\ddot{x} + 2 \lambda \dot{x} + \omega_o^2 x + \alpha \omega_o ^2 x^2 + \beta \dot{x}^2 + \gamma x \ddot{x} = 0.
\label{OscNL}
\end{equation}

To determine the frequency shift induced by a stationary force field, we follow the same approach as presented in section \ref{ss:dom}. We introduce the equilibrium position $x_s+\alpha {x_s}^2=g/{\omega_o}^2$ and then make the substitution $X-x_s$. Equation \ref{OscNL} thus becomes:

\begin{multline*}
\ddot{X} + \frac{2 \lambda}{1 + \gamma (x_s + X)} \dot{X} +\\ \frac{1 + \alpha (2 x_s  + X)}{1 + \gamma (x_s + X)} \omega_o^2 X + \frac{\beta}{1 + \gamma (x_s + X)}   \dot{X}^2 = 0.
\end{multline*}

We then assume that $X \ll x_s$, i.e. the drop oscillation is small compared to the static deformation. This hypothesis is supported by experimental observations on pendant drops (Fig. \ref{fig:hmdh}). Then, by neglecting the influence of the viscosity, we get :

$$
\ddot{X} +  \frac{1 + 2 \alpha x_s }{1 + \gamma x_s} \omega_o^2 X = 0.
$$ 

Finally, for small static deformations ($\gamma x_s, \alpha x_s \ll 1$), the frequency shift induced by a stationary force field is:

$$
\omega_s = \omega_o \left( 1 + (  \alpha - \frac{\gamma}{2}) x_s \right).
$$

Drop frequency shift induced by large oscillations has been theoretically investigated by Tsamopoulos and Brown\cite{Tsamopoulos1983}: they showed that drop frequency evolves as $\omega_d=\omega_o(1+K'A^2)$ with $A$ the amplitude of oscillation, in good agreement with experiments\cite{Trinh1982}.

Now following the Landau method used in section \ref{ss:dom}, the differential equation ruling the time-evolution of $X_2$ is:

\begin{multline*}
\ddot{X_2}+\omega_o^2X_2=\\ \frac{1}{2}\left(\alpha+(\beta-\gamma)\omega_o^2\right)A^2-\frac{1}{2}\left(\alpha-(\beta-\gamma)\omega_o^2\right)A^2\cos(2\omega_d t).
\end{multline*}

By solving this equation, we obtain the following expression for $X_2$ :

$$
X_2=\frac{\left((\gamma-\beta)\omega_o^2-\alpha\right)}{2\omega_o^2}A^2-\frac{\left(\alpha-(\gamma+\beta)\omega_o^2\right)}{2\omega_o^2}A^2\cos(2\omega_d t).
$$

We can then express the mean deformation induced by the nonlinearities of the equation \ref{OscNL}:

$$
x_d=\left< X\right>=\left< X_2 \right>=\frac{\left((\gamma-\beta)\omega_o^2-\alpha\right)}{2\omega_o^2}A^2.
$$

The frequency shift can now be expressed in terms of the average deformation:

$$
\omega_d=\omega_o(1+K'\frac{2\omega_o^2}{\left((\gamma-\beta)\omega_o^2-\alpha\right)}x_d)=\omega_o(1+Kx_d).
$$

Finally, since we perform weakly nonlinear analysis (through asymptotic expansion), the two effects can be added to obtain the drop nonlinear frequency:

$$
\omega_{nl} = \omega_o \left( 1 + (\alpha - \gamma/2) x_s + K x_d \right).
$$


%

\end{document}